# Network robustness assessed within a dual connectivity perspective

Alejandro Tejedor[1*], Anthony Longjas[1], Ilya Zaliapin[3,] Samuel Ambroj[4] , and Efi Foufoula-Georgiou[1,2]

[1]National Center for Earth-surface Dynamics and St. Anthony Falls Laboratory, University of Minnesota, Minneapolis, MN, USA

[2]Department of Civil Engineering, University of Minnesota, Minneapolis, MN, USA

[3]Department of Mathematics and Statistics, University of Nevada, Reno, NV, USA

[4]Steinbuch Centre for Computing, Karlsruhe Institute of Technology (KIT), Karlsruhe, Germany

**Abstract**

Network robustness against attacks has been widely studied in fields as diverse as the Internet, power grids and human societies. Typically, in these studies, robustness is assessed only in terms of the connectivity of the nodes unaffected by the attack. Here we put forward the idea that the connectivity of the affected nodes can play a crucial role in properly evaluating the overall network robustness and its future recovery from the attack. Specifically, we propose a dual perspective approach wherein at any instant in the network evolution under attack, two distinct networks are defined: (i) the Active Network (AN) composed of the unaffected nodes and (ii) the Idle Network (IN) composed of the affected nodes. The proposed robustness metric considers both the efficiency of destroying the AN and the efficiency of building-up the IN. We show, via analysis of both prototype networks and real world data, that trade-offs between the efficiency of Active and Idle network dynamics give rise to surprising crossovers and re-ranking of different attack strategies, pointing to significant implications for decision making.

*corresponding author (alej.tejedor@gmail.com)

Recent developments in understanding the structure and dynamics of networks have transformed research in many fields, ranging from protein interactions in a cell to page connectivity in the World Wide Web and relationships in human societies[1, 2]. Although these complex networks have different evolution rules, many exhibit a universal scale-free topology wherein the highly connected nodes, although sparse, dominate the connectivity of the network[2]. Network robustness against failure and attack has been widely studied, and different strategies to manage perturbation spread within the network have been suggested [3, 4, 5, 6, 7, 8]. The existing works focused mainly on the connectivity of the nodes unaffected by an attack while the connectivity of the affected nodes has received minimal attention. However, it is conceivable that for some processes such as disease or information spreading, river contamination, etc., the dynamic connectivity of the affected nodes (*e.g.*, small vs. large clusters of sick people, or contaminated streams) is of interest too as its structure can exert significant feedbacks to the unaffected nodes, can determine the overall system "health", or can establish the propensity of the system to future perturbations. In this paper, we highlight the importance of incorporating information from both the Active Network (unaffected nodes) and Idle Network (affected nodes) into making assessments of the network robustness and evaluating possible interventions in response to an attack.

Consider a network $N$ that consists of nodes $\{n_i\}$, $i = 1,…,T$ connected by edges $\{(n_i,n_j)\}$. We focus on a process of *sequential node removal*, also called an *attack*. The process starts at $t = 0$ with the original network $N$. At each discrete time step $t > 0$ it eliminates a suitably chosen node $n_i$ and all edges $(n_i,\bullet)$ connected to this node, resulting in the set of nodes and edges that have been unaffected and thus are active at $t$, called the Active Network $N_A(t)$. This sequential node removal can mimic a multitude of actual processes operating on networks and having a binary outcome, *e.g.* healthy species in a biological community that may become sick, clean streams in a river network that may become contaminated, people that may learn particular information, *etc*. We also consider the Idle Network $N_I(t)$ that consists of the nodes that have been removed from $N$ up to time $t$, together with all the edges from $N$ among these idle nodes. Accordingly, a sequential node removal process $D$ results in the following decomposition of the network $N$:

$$D: N \rightarrow \{N_A(t), N_I(t)\}, t = 1,\ldots,T. \qquad (1)$$

Observe that the union of the nodes in the $N_A(t)$ and $N_I(t)$ matches the set of nodes in the original network $N$. At the same time, the union of edges from $N_A(t)$ and $N_I(t)$ is only a subset of the edges in the original network, since the latter may also include some edges between nodes in the $N_A(t)$ and $N_I(t)$ representing the possible interactions between the AN and IN. In other words, the pair $\{N_A(t), N_I(t)\}$ cannot be used in general to reconstruct $N$; although $N_A(t)$ is uniquely determined by $\{N, N_I(t)\}$ and $N_I(t)$ is uniquely determined by $\{N, N_A(t)\}$.

The existing literature mainly focuses on studying connectivity metrics on $N_A(t)$ to evaluate the robustness of the network $N$, *i.e.*, the ability of $N$ to preserve connectivity and thus functionality under attack. We assert that a robustness metric of the network $N$ should consider both the dynamics of $N_A(t)$ and $N_I(t)$. We begin by illustrating the importance of this dual perspective by considering an example of node removal in a simple line-connected network of length $T = 7$ shown in Fig. 1. The connectivity of a network is assessed here by the size $S(t)$ of its largest cluster; this is a conventional metric used in many previous studies [3, 4, 5, 6, 7, 8]. We implemented a strategy of node removal that is the most efficient in decreasing the size $S_A(t)$ of the maximal cluster of the AN (Fig. 1a). During the first three time steps the max cluster size decreased from 7 to 1. However, this particular strategy of node removal is not at all efficient with respect to building-up the connectivity of the IN (Fig. 1b): in the first three time steps the maximum cluster size $S_I(t)$ merely increased from 0 to 1.

Quantitatively, the *efficiency* $E_A$ of a node removal strategy in destroying the AN can be defined as:

$$E_A = \frac{A_A}{A_{max}} = \frac{\sum_{t=0}^{T}(T-t-S_A(t))}{\sum_{t=0}^{T}(T-t)} = 1 - \frac{2}{T(T+1)}\sum_{t=0}^{T} S_A(t). \quad (2)$$

Here $A_A$ is the area between $S_A(t)$ and the diagonal staircase $(T-t)$ as in Fig. 1c and $A_{max}$ is the area below the diagonal staircase. Similarly, the efficiency $E_I$ of building the idle network can be defined as (Fig. 1d):

$$E_I = \frac{A_I}{A_{max}} = \frac{\sum_{t=0}^{T} S_I(t)}{\sum_{t=0}^{T}(T-t)} = \frac{2}{T(T+1)}\sum_{t=0}^{T} S_I(t) \quad (3)$$

We propose to define the network robustness $R_N$ as a function of both the efficiency $E_A$ of destroying the connectivity of the AN and efficiency $E_I$ of building-up the connectivity of the IN:

$$R_N = f(E_A, E_I) \quad (4)$$

with a suitable function $f$ non-increasing in both arguments. In the absence of specific reasons for non-linearity, a simple metric of network robustness would be:

$$R_N(\alpha) = \alpha(1 - E_A) + (1 - \alpha)(1 - E_I), \quad (5)$$

where $\alpha$ is the weight given to the efficiency of the AN while $(1-\alpha)$ is the complementary weight given to that of the IN. Using $\alpha = 1$ leads to a particular definition that is currently used in the literature to guide, for example, decisions on most effective strategies of attack or to assess recovery rates under a given attack [3, 4, 5, 6, 7, 8]. While this may be a good approximation for some systems, it is restrictive for many others. For example, the robustness of the Internet has been studied under different attacks[4,5], wherein the routers are the nodes of the network, and the wire or wireless connections are the edges. The robustness of these systems to withstand an attack has been assessed by considering only the connectivity of the unaffected routers, *i.e.*, the sooner the network under attack losses connectivity, the less robust it is. This hypothesis is considering only one of the perspectives, *i.e.*, connectivity of the AN, to assess the robustness of the overall network ($\alpha = 1$). However, relevant information is disregarded: different scenarios of the connectivity in the IN for the same connectivity in the AN are possible and this can result in different "effective" overall system robustness. For instance, if the failed routers are scattered (low $S_I$) compared to being clustered on specific parts of the network (high $S_I$), different $\alpha < 1$ values could be considered to capture possible trade-offs on the relative importance of the connectivity of the AN and IN in assessing the overall system robustness to the attack. Similar examples apply to balancing the spread of one ecological species at the expense of another, containment of contaminated waters in water corridors or in spreading of diseases and information. The implications of the above trade-offs for decision-making are apparent.

To illustrate some subtle and unexpected consequences that arise in considering a dual perspective in defining network robustness under an attack, three types of networks and three different strategies of node elimination (attack strategies) are studied. Network 1: A *square lattice* of $T = 10,000$ nodes arranged in a Von Neumann neighborhood (*i.e.*, each node having four neighbors); see Fig. 2a. Network 2: A *Tokunaga self-similar tree*[9] (T-tree) with parameters (*a, c*) = (1, 2) (see Fig. 2e). Network 3: A Barabasi-Albert (BA) *scale-free network*, a system with heterogeneous node degree distribution that exhibits high connectivity and contains intricate structures due to the presence of loops. The Tokunaga self-similar tree is known to describe a critical binary Galton-Watson process[10] and level-set tree representation of a symmetric random walk or regular Brownian motion[11]. Tokunaga trees with a broad range of parameter values have found wide applicability in describing the topology of river networks[9, 12, 13], biological networks (leaves and cardiovascular systems) [14] and clustering of earthquake aftershocks[15]. In this paper we use Tokunaga trees of order $\Omega = 6$. Each Horton-Strahler branch[12, 16] in the tree represents a node. The BA network incorporates preferential attachment and growth mechanisms[2]. We construct a BA network using an initially connected network of $m_0 = 3$ nodes and adding a new node with $m = 2$ links per time step, until $T = 1,000$ nodes are added (Fig. 2i). The examined networks are classified according to the node degree distribution into *homogeneous* (lattice) and *heterogeneous* (T-tree and BA network).

In each system, we examine three strategies of node removal. Strategy 1: A *random failure* (RF) removes nodes at random using a discrete uniform distribution over all the active nodes. Strategy 2: A *targeted attack* (TA) assigns a removal probability to a vertex proportional to its degree of connectivity in the AN. Strategy 3: A *random spreading* (RS) removes the first node at random as in RF; afterwards, at each time step one node connected to an eliminated node is randomly removed. The evolution of the largest cluster size $S$ under progressive node removal is examined using 100 simulations. Figure 2 shows $S(t)$ as a function of time in the AN and IN for one realization (representative of all simulations) for each network and attack; the time $t$ is normalized to be equal to the fraction of the removed nodes. The first observation is that the rate of increase of the largest cluster size in the IN is not the same, in general, as the rate of decay of the largest cluster size in the AN. A lattice network under RS is an exception – the symmetry here (with respect to $S(t) = 0.5$) is expected by construction and it can only be altered by abrupt jumps in $S(t)$ due to finite size effects. We also notice symmetry of $S_A(t)$ and $S_I(t)$ with respect to the vertical axis $t = 0.5$ that is only observed for a homogeneous lattice network (under any attack) and random failure (applied to any network) and can be expressed as $E_A + E_I \approx 1$ (see Table 1). The symmetry is not obvious in Fig. 2f due to the large jumps of the largest cluster size; although it can be shown statistically via the efficiency values (Table 1). Having the complementary values of $E_A$ and $E_I$ has an obvious but important implication: the more efficient a strategy according to one perspective (*e.g.*, destroying the connectivity in the AN), the less efficient it is according to the other (*e.g.*, building-up the connectivity in the IN). Another important observation is that for T-trees, the connectivity of the AN is destroyed faster, and the connectivity of the IN is built up slower, than in the BA network. Finally, the perfect efficiency of the random spreading in the Idle network is a consequence of its definition ($S_I$ grows linearly).

The robustness (equation 5) of a network does change with $\alpha$, as illustrated in Fig. 3 (top panels). Notably, the robustness may deviate substantially from the case $\alpha = 1$ (marked by stars in Fig. 3), which is examined in most of the existing studies[3, 4, 5, 6, 7, 8]. Surprisingly, a more general definition (equation 5) not only gives different numerical values of the robustness, but also may result in *robustness crossovers* – alternative ranking of attack strategies depending on the value of $\alpha$. For example, in a lattice network, a crossover occurs at $\alpha = 0.5$, with $R_{N,RS} > R_{N,RF} > R_{N,TA}$ for $\alpha > 0.5$, and $R_{N,TA} > R_{N,RF} > R_{N,RS}$ for $\alpha < 0.5$ (here the second lower index refers to the attack type). A crossover between $R_{N,TA}$ and $R_{N,RS}$ is also observed for T-trees at $\alpha \approx 0.68$ as well as for the BA network at $\alpha \approx 0.17$. Hence, an interplay between the AN and IN introduces a whole new dimension in the study of robustness, which cannot be reproduced by exclusively examining the AN. At the same time, some general observations remain consistent with previous works when $\alpha = 1$, in particular those showing that networks are more robust under random failure than targeted attack [3, 5, 7]. Other observations for $\alpha = 1$ are: (1) for both the heterogeneous networks, $R_N$ is highest for random failure, followed by random spreading and targeted attack; (2) the robustness in homogeneous networks is highest for random spreading, followed by random failure and targeted attack; (3) the $R_N$-value for random spreading and homogeneous networks is approximately equal to 1 since $S_I$ grows linearly by definition and the efficiencies are complementary ($E_A = 0$, $E_I = 1$).

The results presented so far considered that the same node removal rules (time-invariant attack) operated on the system until its complete destruction. In many systems however, an adoptive "attack and recovery strategy" is applied, *i.e.*, system performance is evaluated periodically, and especially in the early stages of the attack, to guide future actions. It is understood, for example, that an attack strategy, which is optimal when evaluated over a long period of time might be suboptimal relative to a shorter time horizon. Figure 3 (bottom panels) shows the results of the robustness-based ranking of attack strategies defined with respect to a partial (10%) system destruction. Although both the strong dependence of robustness on $\alpha$ and the presence of crossovers is still observed, the crossover location moves closer to $\alpha = 1$ with substantial divergence in the attack strategy rankings for $\alpha < 1$. The practical implications of this finding can be substantial; for example in a BA network $\alpha = 0.7$ (which gives 70% weight to the AN and 30% to the IN) would remarkably re-rank the robustness of different attack strategies which for $\alpha = 1$ would be indistinguishable (rightmost bottom panel plot of Fig. 3).

To further illustrate the importance of the Idle Network in assessing system robustness, we consider data from the second largest European airline, RyanAir[17]. The examined network consists of 186 airports and 1507 edges that represent the existence of at least one weekly flight between the respective airports (Figure 4a). Figure 4b shows the robustness values for a sequential removal of airports until all of them are inoperative (100% removal), according to the three previously implemented attack strategies: RF, TA and RS. The results are consistent with our simulations: (1) Random Failure generates complementary efficiencies ($E_A+E_I \approx 1$). (2) Robustness ranking for the different strategies is similar to the BA network (see Figure 3) and the crossover between TA and RS is observed near α = 0. Figure 4c shows the network robustness when the attack strategies act only until 19 airports are removed (10% node removal). Qualitatively, we have the same behavior as in the BA network (cf. Figure 3). However there exist significant quantitative differences, expressed in much lower values of robustness for $\alpha < 1$. This is due to the structure of the airline network (point-to-point), which has numerous connections among all the airports and hence relatively high connectivity degree for all nodes, not only the hubs. Thus, it is more likely to build-up clusters in the IN than hub-and-spoke scale-free networks.

To further illustrate why a value of $\alpha < 1$ might be imperative to consider, in Fig 5 we display two transportation networks whose operative airport (AN) connectivity is indistinguishable under two different attacks but that of their inoperative airports (IN) is drastically different. Since the largest cluster size in the Active network under both attacks is the same, $S_A = 167$ airports, the standard metric of robustness ($\alpha = 1$) would rank them equally robust (see also Fig. 5c). However, the largest cluster size in the IN of Fig. 5a is more than 6 times bigger compared to the one shown in Figure 5b. It is obvious that due to economic, logistic, security, and other aspects, both scenarios are significantly different. Consider for instance the monetary losses, conceptually approximated by the amount of traffic lost due to removal of airports (temporarily due to weather, or permanently due to structural airline reorganization). We roughly approximate the lost traffic by the number of lost edges in the network, and use the largest cluster size in IN as a proxy for this quantity. Naturally, losing a certain number of well-connected airports leads to more severe traffic losses than losing the same number of disconnected airports. This difference is picked up by our robustness measure for $\alpha < 1$. This example illustrates the necessity of incorporating a dual perspective framework in evaluating the overall robustness of a system under an attack by giving weight to the Idle Network, which has been systematically ignored in literature.

**Acknowledgments**

We acknowledge Hawoong Jeong and Jonathan Czuba for useful discussions. The research is supported by NSF grant EAR-1209402 under the Water Sustainability and Climate Program.

Table 1 **Efficiencies for the three attack strategies applied to the lattice, T-tree and BA network**

| Attack | Lattice | T-Tree | BA Network |
|---|---|---|---|
| Random Failure (RF) | **$E_A = 0.35 \pm 0.01$**<br>**$E_I = 0.65 \pm 0.01$** | $E_A = 0.58 \pm 0.10$<br>$E_I = 0.39 \pm 0.10$ | **$E_A = 0.17 \pm 0.02$**<br>**$E_I = 0.83 \pm 0.02$** |
| Targeted Attack (TA) | **$E_A = 0.42 \pm 0.00$**<br>**$E_I = 0.58 \pm 0.01$** | $E_A = 0.87 \pm 0.04$<br>$E_I = 0.75 \pm 0.04$ | $E_A = 0.48 \pm 0.02$<br>$E_I = 0.94 \pm 0.01$ |
| Random Spreading (RS) | **$E_A = 0.02 \pm 0.02$**<br>**$E_I = 1$** | $E_A = 0.76 \pm 0.09$<br>$E_I = 1$ | $E_A = 0.19 \pm 0.02$<br>$E_I = 1$ |

$E_A$ ($E_I$) is the efficiency of an attack strategy in destroying (building) the Active (Idle) network.
Values in bold represent complementary efficiencies ($E_A + E_I \approx 1$).

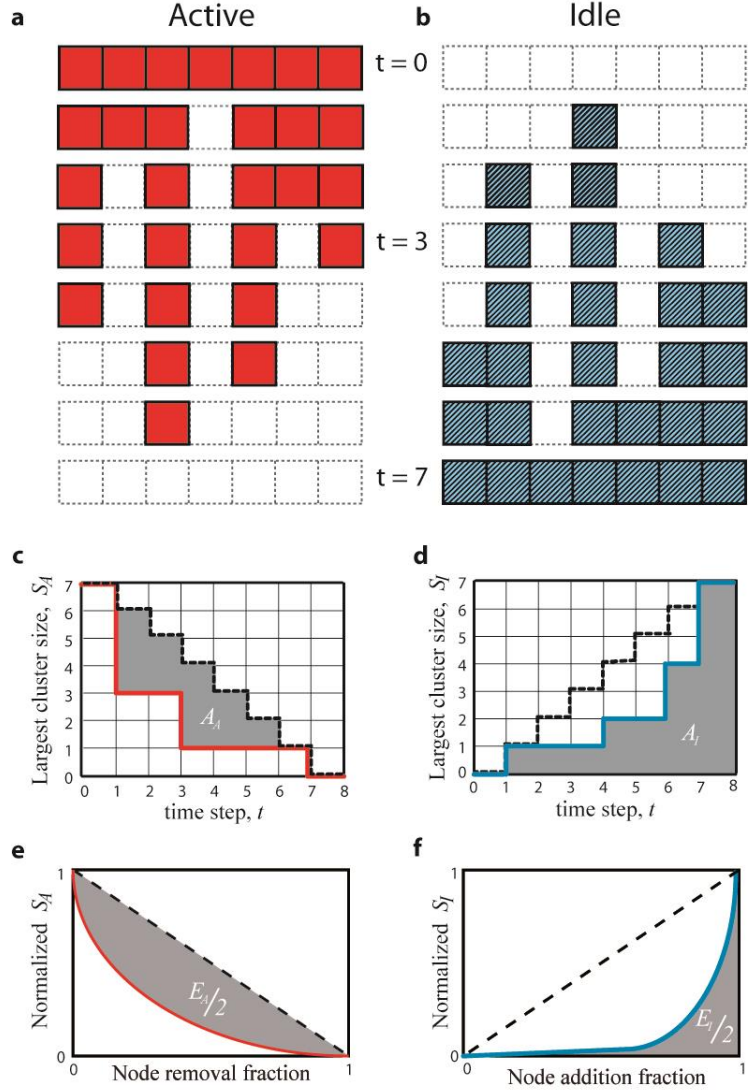

**Figure 1** | **Dual connectivity perspective in a simple line network. a,** At time $t = 0$ the line network consists of seven nodes, all belonging to the Active Network (AN) shown as solid squares. At each time step, one node is removed to destroy the connectivity of the AN in the most efficient way. **b,** Each removed node in the AN creates a node in the Idle Network (IN) shown as stripped squares. The largest cluster size $S_A$ ($S_I$) in the AN (IN) is shown by a solid line in panel **c** (**d**). It is observed that $S_A$ and $S_I$, evolve asymmetrically: the most efficient procedure to reduce $S_A$ is not the most efficient to increase $S_I$. The efficiency of an attack has two components, $E_A$ and $E_I$, one for each perspective, and their values are proportional to the gray area in panels **e** and **f** respectively. This illustrates that defining robustness in terms of only efficiency $E_A$ or in terms of both efficiencies $E_A$ and $E_I$ could make a significant difference in assessing the overall system robustness.

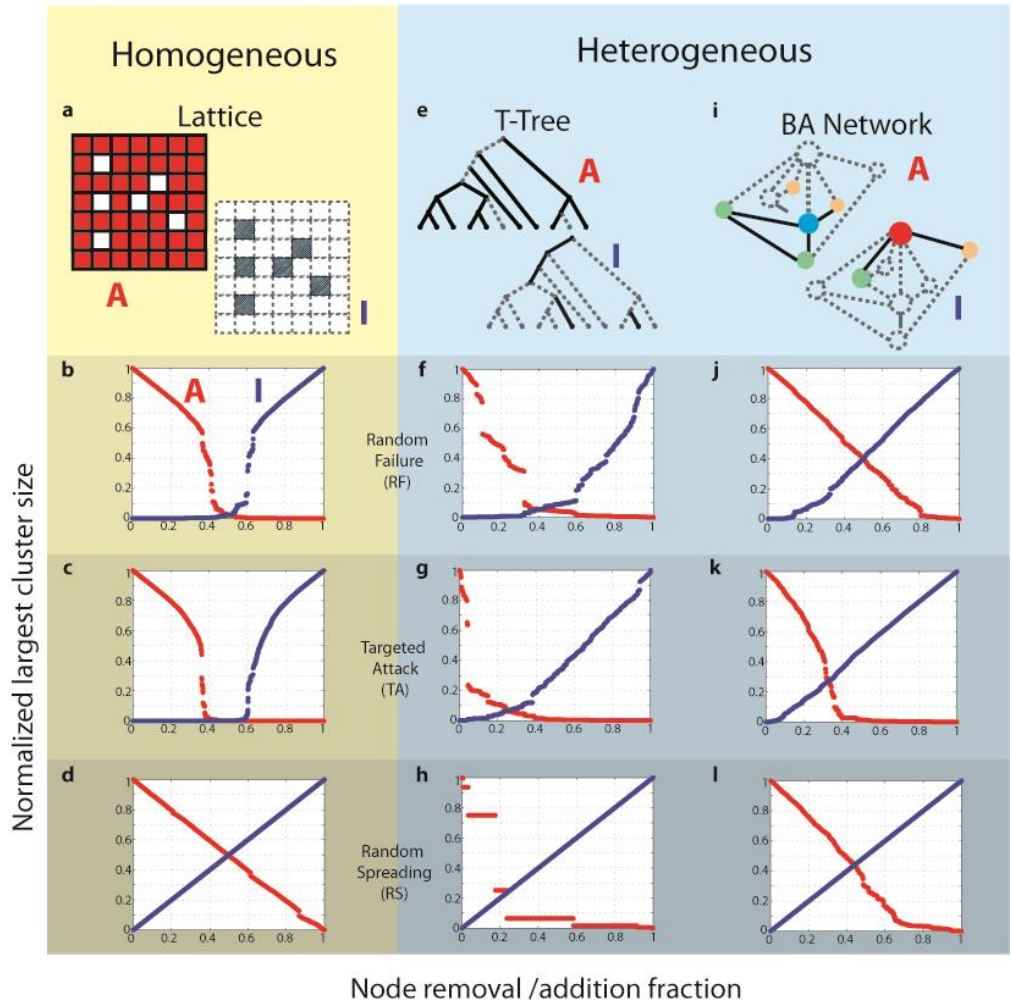

**Figure 2 | Dual perspective evolution of networks under attack.** Evolution of the largest cluster size in the Active Network, AN (red) and Idle Network, IN (blue) for homogeneous (yellow panels) and heterogeneous (blue panels) networks with respect to three different sequential node removal strategies: panels **b, f, j** – random failure, **c, g, k** – targeted attack, and **d**, **h**, **l** – random spreading. The largest cluster size and time are normalized by the system size. Three main observations are made: (i) the rate of decrease of the largest cluster size in the AN is not the same as the rate of increase of the largest cluster size in IN (asymmetric evolution); (ii) for homogenous networks and networks under random failure, there is a symmetry with respect to the vertical axis at 0.5 implying a complementarity in the efficiencies of destroying AN and building-up IN, *i.e.* $E_A + E_I \approx 1$; and (iii) for heterogeneous networks (T-Trees and BA networks) and heterogeneous attacks (TA and RS) no symmetry is observed at all, there is a necessity to monitor both networks (AN and IN) since it is not possible to predict the value of the efficiency of building-up the IN from the efficiency value of destroying the AN and vice versa.

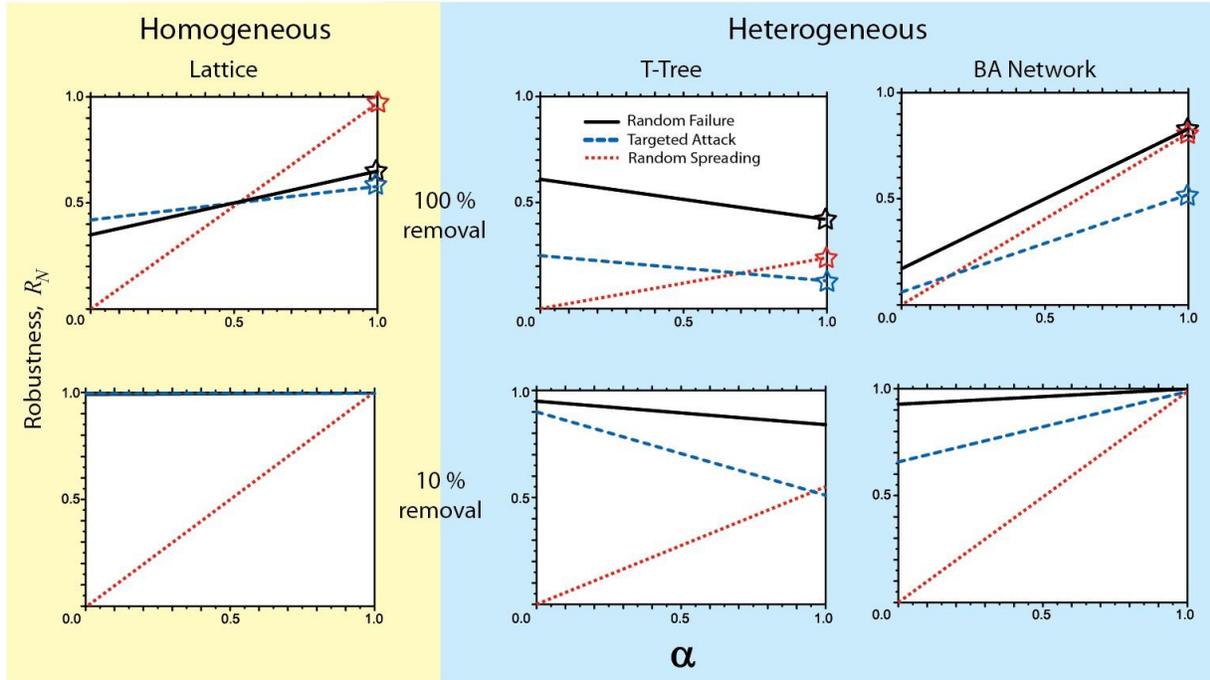

**Figure 3 | Robustness, $R_N$, as a function of the relative weight given to the connectivity of the Active Network (AN), $\alpha$.** The robustness defined exclusively in terms of the AN ($\alpha = 1$) is shown by stars. For the top panels, the robustness of a homogeneous network subject to any attack and heterogeneous networks under random failure, is equal to 0.5 for $\alpha = 0.5$ due to the property $E_A + E_I \approx 1$. For all cases, notice (i) a strong *dependence* of robustness on $\alpha$, (ii) *robustness crossovers* – changes in ranking (ordering of respective $R_N$ values) of different attack strategies depending on $\alpha$ and (iii) shift of the *robustness crossovers* towards $\alpha = 1$ with substantial divergence in the attacks strategies when the system is evaluated not at the time of complete destruction but at its early stages of attack.

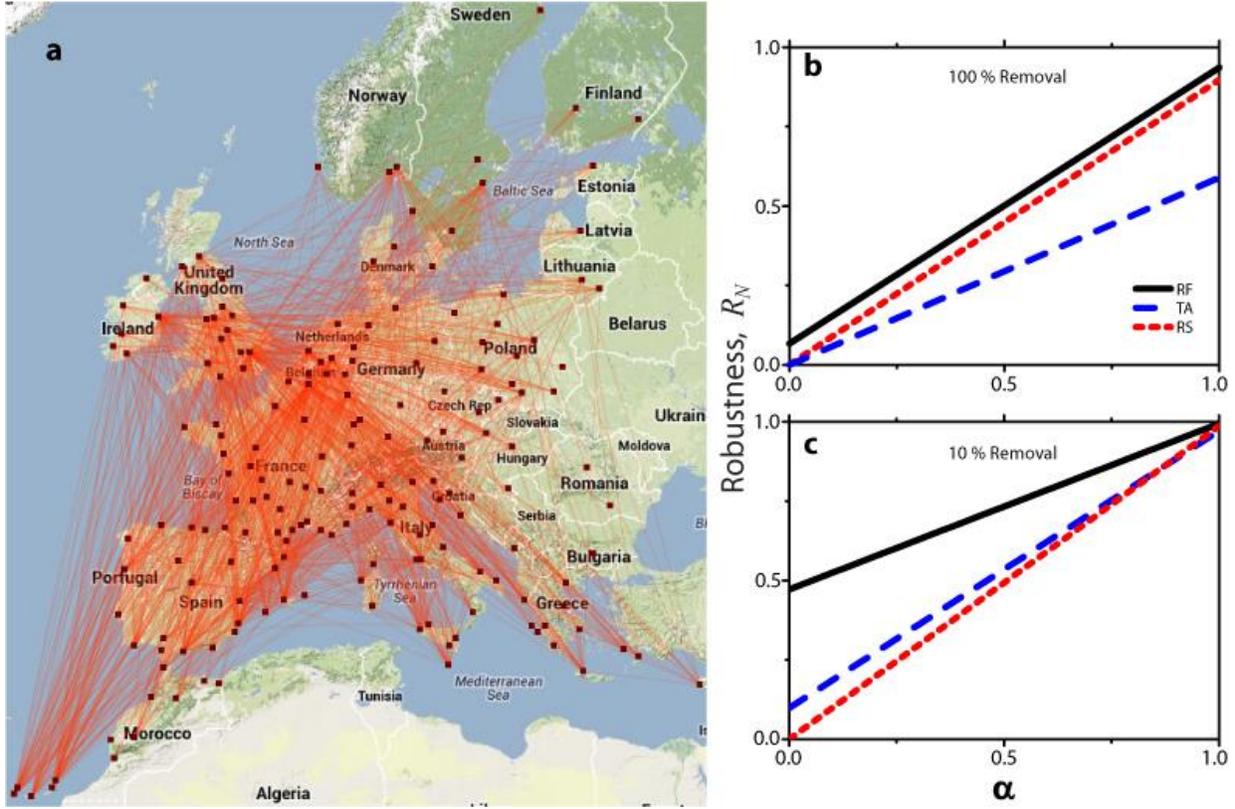

**Figure 4 | Robustness, $R_N$, of the second largest airline network in Europe. a**, The map shows the connectivity of the second largest airline in Europe (RyanAir), operating in 186 airports (black squares), with more than 1500 routes (red lines). **b**, shows the robustness of that network under three different attacks (Random Failure, Targeted Attack and Random Spreading), which act until all the airports are removed. In **c**, the robustness is evaluated under partial attack (10 % airports removed). Note that for $\alpha = 1$, the network is equally robust under any of the three attacks. However, for $\alpha < 1$ the robustness values for different attack strategies significantly differ, highlighting the importance of the Idle Network in assessing the robustness of the system.

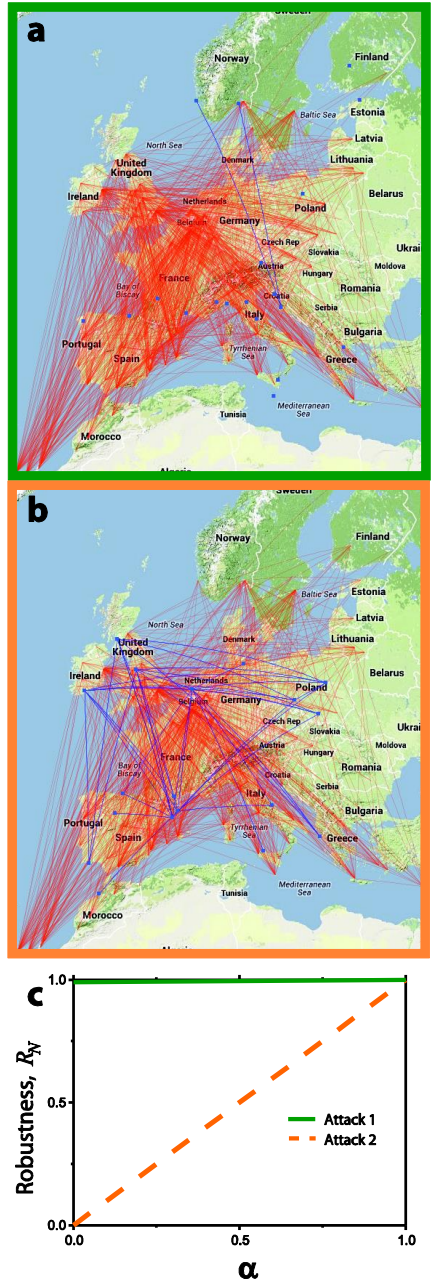

**Figure 5 | Importance of the dual perspective framework in assessing network robustness.** The maps illustrate the result of two different attacks, attack 1 and attack 2, applied to the network until 19 airports are disconnected (10% removal). The two resulting networks have the same largest cluster size in the Active Network ($S_A = 167$), but different largest cluster size in the Idle Network: **a,** $S_I = 3$ for attack 1 and **b,** $S_I = 19$ for attack 2. Considering connectivity of the Idle Network in assessing the system robustness reveals significant differences in these two attacks, as quantified in **c**.

17